\documentclass[12pt]{article}

\begin{document}

\title{\textbf{Comment on ``Fun and frustration with quarkonium in a 1+1
dimension,'' by R. S. Bhalerao and B. Ram [Am. J. Phys. 69 (7), 817-818
(2001)]}}
\date{}
\author{Antonio S. de Castro\thanks{%
Electronic mail: castro@feg.unesp.br} \\
UNESP/Campus de Guaratinguet\'{a}\\
Caixa Postal 205\\
12500-000 Guaratinguet\'{a} SP, Brasil}
\maketitle

\baselineskip=20pt

\newpage

In a recent article to this journal, Bhalerao and Ram \cite{ram} approached
the Dirac equation in a 1+1 dimension with the scalar potential

\begin{equation}
V(x)=g|x|  \label{eq0}
\end{equation}

\noindent For $x>0$ the Dirac equation

\begin{equation}
\left[ \left( 
\begin{array}{cc}
1 & 0 \\ 
0 & -1
\end{array}
\right) \frac{d}{dx}+\left( m+gx\right) \right] \psi =\left( 
\begin{array}{cc}
0 & 1 \\ 
1 & 0
\end{array}
\right) E\psi  \label{eq01}
\end{equation}

\noindent reduces to

\begin{eqnarray}
\left( -\frac{d^{2}}{d\xi ^{2}}+\xi ^{2}\right) \psi _{1} &=&\left( \frac{%
E^{2}}{g}+1\right) \psi _{1}  \nonumber \\
&&  \label{eq1} \\
\left( -\frac{d^{2}}{d\xi ^{2}}+\xi ^{2}\right) \psi _{2} &=&\left( \frac{%
E^{2}}{g}-1\right) \psi _{2}  \nonumber
\end{eqnarray}

\noindent with $\xi =\sqrt{g}\left( m/g+x\right) $. $\psi _{1}$ and $\psi
_{2}$ are the upper and the lower components of bispinor $\psi $,
respectively. For $x<0$ the equations take the form

\begin{eqnarray}
\left( -\frac{d^{2}}{d\xi ^{\prime 2}}+\xi ^{\prime 2}\right) \psi _{1}
&=&\left( \frac{E^{2}}{g}-1\right) \psi _{1}  \nonumber \\
&&  \label{eq2} \\
\left( -\frac{d^{2}}{d\xi ^{\prime 2}}+\xi ^{\prime 2}\right) \psi _{2}
&=&\left( \frac{E^{2}}{g}+1\right) \psi _{2}  \nonumber
\end{eqnarray}

\smallskip \noindent with $\xi ^{\prime }=\sqrt{g}\left( m/g-x\right) $. The
authors of Ref. [1] state that the solutions for Eq.(\ref{eq1}) are

\begin{eqnarray}
\psi _{1} &=&CH_{n+1}(\xi )e^{-\xi ^{2}/2}  \nonumber \\
&&  \label{eq3} \\
\psi _{2} &=&DH_{n}(\xi )e^{-\xi ^{2}/2}  \nonumber
\end{eqnarray}

\noindent where $H_{n}(\xi )$ are the Hermite polynomials. Similar solutions
are found for $x<0$. Substituting (\ref{eq3}) into the Dirac equation they
found

\begin{equation}
E=\pm \sqrt{2\left( n+1\right) g}  \label{eq4}
\end{equation}

\noindent whereas the continuity of the wave function at $x=0$ leads to the
quantization condition

\begin{equation}
H_{n+1}(\alpha )=\pm \left[ 2\left( n+1\right) \right] ^{1/2}H_{n}(\alpha )
\label{eq5}
\end{equation}

\noindent with

\begin{equation}
\alpha =\frac{m}{\sqrt{g}}  \label{eq5.1}
\end{equation}
\noindent The authors regretted they found only the ground-state solution
for $\alpha =1/\sqrt{2}$. No solutions were found with numerical evaluation
of $n$ from 1 to 250. Hence the authors concluded that \textit{it is not
possible to describe the meson spectrum with (\ref{eq0}) as the
quark-antiquark potential when used as a Lorentz scalar in the 1+1
-dimensional Dirac equation.}

The equations (\ref{eq1})-(\ref{eq2}) can also be cast into the form

\begin{eqnarray}
\frac{d^{2}\psi _{1}}{d\eta ^{2}}-\left( \frac{\eta ^{2}}{4}-\nu -\frac{3}{2}%
\right) \psi _{1} &=&0  \nonumber \\
&&  \label{eq6} \\
\frac{d^{2}\psi _{2}}{d\eta ^{2}}-\left( \frac{\eta ^{2}}{4}-\nu -\frac{1}{2}%
\right) \psi _{2} &=&0  \nonumber
\end{eqnarray}

\smallskip 
\begin{eqnarray}
\frac{d^{2}\psi _{1}}{d\eta ^{\prime 2}}-\left( \frac{\eta ^{^{\prime }2}}{4}%
-\nu -\frac{1}{2}\right) \psi _{1} &=&0  \nonumber \\
&&  \label{eq7} \\
\frac{d^{2}\psi _{2}}{d\eta ^{\prime 2}}-\left( \frac{\eta ^{^{\prime }2}}{4}%
-\nu -\frac{3}{2}\right) \psi _{2} &=&0  \nonumber
\end{eqnarray}

\noindent where $\eta =\sqrt{2}\xi $, $\eta ^{\prime }=\sqrt{2}\xi ^{\prime
} $ and

\begin{equation}
\frac{E^{2}}{2g}=\nu +1  \label{eq8}
\end{equation}

The second-order differential equations (\ref{eq6})-(\ref{eq7}) have the form

\begin{equation}
y^{^{\prime \prime }}(z)-\left( \frac{z^{2}}{4}+a\right) y(z)=0  \label{eq9}
\end{equation}

\noindent whose solution is called a parabolic cylinder function \cite{as}.
The solutions $D_{-a-1/2}(z)$ and $D_{-a-1/2}(-z)$ are linearly independent
unless $n=-a-1/2$ is a nonnegative integer. In that special circumstance $%
D_{n}(z)$ has the peculiar property that $D_{\nu }(-z)=(-1)^{n}D_{\nu }(z)$
and it is proportional to $\exp \left( -z^{2}/4\right) H_{n}(z/\sqrt{2})$,
in which $H_{n}(z)$ is a Hermite polynomial. In the circumstance in hands
the solutions do not exhibit such parity properties, instead they are

\begin{eqnarray}
\psi (x &>&0)=\left( 
\begin{array}{c}
CD_{\nu +1}(\eta ) \\ 
DD_{\nu }(\eta )
\end{array}
\right)  \nonumber \\
&&  \label{eq10} \\
\psi (x &<&0)=\left( 
\begin{array}{c}
C^{\prime }D_{\nu }(\eta ^{\prime }) \\ 
D^{\prime }D_{\nu +1}(\eta ^{\prime })
\end{array}
\right)  \nonumber
\end{eqnarray}

\noindent When these solutions are inserted into the Dirac equation one
obtains

\begin{equation}
\frac{D}{C}=\frac{C^{\prime }}{D^{\prime }}=\frac{E}{\sqrt{2g}}  \label{eq11}
\end{equation}

\noindent In the last steps the recurrence formula

\begin{equation}
D_{\nu }^{\prime }(z)-z/2D_{\nu }(z)+D_{\nu +1}(z)=0  \label{eq11.1}
\end{equation}

\noindent has been used for obtaining $D_{\nu }^{\prime }(z)$. On the other
hand, the joining condition at $x=0$ leads to the quantization condition

\begin{equation}
D_{\nu +1}\left( \sqrt{2}\alpha \right) =\pm \sqrt{\nu +1}D_{\nu }\left( 
\sqrt{2}\alpha \right)  \label{eq12}
\end{equation}

\noindent where $\alpha $ is given by (\ref{eq5.1}).

Our derivation of the quantization condition does not involve the
restriction that $n$ is an integer number and that should not be done since
the differential equations are not the same for different segments of the
axis $X$. The numerical computation of (\ref{eq12}) is substantially simpler
when $D_{\nu +1}\left( \sqrt{2}\alpha \right) $ is written in terms of $%
D_{\nu }^{\prime }\left( \sqrt{2}\alpha \right) $:

\begin{equation}
D_{\nu }^{\prime }\left( \sqrt{2}\alpha \right) =\left[ \frac{\alpha }{\sqrt{%
2}}\mp \sqrt{\nu +1}\right] D_{\nu }\left( \sqrt{2}\alpha \right)
\label{eq13}
\end{equation}

\noindent With $\alpha =1/\sqrt{2}$ and arbitrarily choosing $D_{\nu }\left( 
\sqrt{2}\alpha \right) $one finds that $\nu $ for the lowest states is given
by 2.042$\times 10^{-6}$ and 2.681, for the minus signal in (\ref{eq13}). On
the other hand the plus sign gives 1.524 and 3.914.

\vspace{5cm}

\textbf{Acknowledgments}

Work supported in part through funds provided by CNPq and FAPESP.

\vfill\eject

\end{document}